%% file: R66636-1a.tex
\documentclass[preprint,showpacs,preprintnumbers,amsmath,amssymb]{revtex4}

\usepackage{graphicx}
\usepackage{dcolumn}
\usepackage{bm}

\begin{document}

\preprint{APS/123-QED}
\input{commands.tex}

\title{Numerical Study of TAP Metastable States in 3-body Ising Spin Glasses}

\author{Yukinori Tonosaki}
\author{Koujin Takeda}%
\author{Yoshiyuki Kabashima}
\affiliation{Department of Computational 
Intelligence and Systems Science, 
Tokyo Institute of Technology, Yokohama 2268502, Japan}

\date{\today}

\begin{abstract}
The distribution of solutions of the Thouless-Anderson-Palmer equation is studied by extensive numerical experiments for fully connected $3$-body interaction Ising 
spin glass models in a level of annealed calculation. A recent study predicted that when the equilibrium state of the system is characterized by one-step replica symmetry breaking, the distribution is described by a Becchi-Rouet-Stora-Tyutin (BRST) supersymmetric solution in the relatively low free energy region, whereas the BRST supersymmetry is broken for higher values of free energy (Crisanti et al., Phys. Rev. B \textbf{71} (2005) 094202). Our experiments qualitatively reproduce the discriminative behavior of macroscopic variables predicted by the theoretical assessment. 
\end{abstract}

\maketitle

\section{Introduction} 
The Thouless-Anderson-Palmer (TAP) approach is a representative method for investigating disordered systems of mean field type \cite{TAP1978}. Unlike the replica method, which is another method employed in the research of disordered systems 
for examining average behavior, this scheme makes it possible to analyze a system 
realized by a specific sample of quenched disorder. Using an appropriate approximation of mean field type, the TAP approach yields a set of coupled nonlinear equations, which is termed the TAP equation. Solutions of the TAP equation provide
approximate averages of state variables for the specific quenched disorder, which are expected to provide accurate values in the limit of the infinite system size when adequate treatment is provided. 

For many systems, the TAP equation can have multiple solutions, the number of which is expected to grow exponentially with respect to the systems size when frustration becomes sufficiently strong in low temperatures. This indicates that exponentially many metastable states can exist for an approximate free energy, referred to as the TAP free energy, and is considered to be related to the origin of the replica symmetry breaking (RSB). Therefore, analytically evaluating the number of metastable states is important. 

More than a quarter of a century has passed since the first analytical 
estimate of the number of metastable states was performed 
by Bray and Moore (BM) \cite{BrayMoore1980} 
for the Sherrington-Kirkpatrick (SK) model \cite{SK1975}. 
However, with the recent introduction of the concept of Becchi-Rouet-Stora-Tyutin (BRST) supersymmetry (SUSY) from quantum field theory, such assessments have attracted a great deal of attention \cite{Cavagna2003}. This concept shed new light on the solution of the BRST SUSY type, which previously attracted little attention, and on the solution of BM, for which the BRST SUSY is broken. 
A recent theoretical study indicated that the BRST SUSY solution 
is never relevant for the SK model in the level 
of annealed assessment \cite{Crisanti2003}, 
as verified by extensive numerical experiments \cite{Cavagna2004}. 
Unfortunately, solving the TAP equation is technically difficult 
in general \cite{TAPdifficulty}. Therefore, as far as the authors know, 
numerical validation of the theoretical results has yet been not sufficiently performed, particularly for $p$-body interaction systems, which are considered
to exhibit an RSB that differs from that of the SK model. 

Thus, we herein investigate the $p$-body Ising spin models of mean field type numerically, in particular, for $p=3$. More precisely, we perform extensive numerical experiments for TAP equations of $3$-body systems and compare the results with the theoretical predictions for the level of the annealed assessment, which has been obtained in a previous study \cite{Crisanti2005}. The previous investigation utilizing the BRST SUSY concept indicates that the distribution of the metastable states of $p (\ge 3)$ systems in the RSB regime is described by the BRST SUSY and BRST SUSY breaking solutions for low and high free energy regions, respectively, whereas only the BRST SUSY breaking solution is relevant in the case of $p=2$. The main objective of the present study is to numerically verify the relevance of the BRST SUSY solution of an RSB phase for $p\ge 3$ systems. 

The present paper is organized as follows. In the next section, we briefly review how the BRST SUSY concept is introduced in counting the number of TAP metastable states and introduce the model considered herein. In Section 3, the scheme of the numerical experiments is explained and the obtained results are compared with the analytical predictions. The final section is devoted to a summary. 

\section{TAP Complexity and Analytical Evaluation}
\subsection{Model and TAP equation}
Although the numerical study is performed only for $3$-body systems, we describe the general model for $p$-body systems because the results obtained from the experiments are considered to be qualitatively relevant for general cases in which $p \ge 3$. The $p$-body spin glass model is defined by the following Hamiltonian:
\begin{eqnarray}
H(\bS)=-\sum_{(i_1 i_2 \ldots i_p ) }
J_{i_1 i_2 \ldots i_p } S_{i_1} S_{i_2} \ldots S_{i_p}, 
\label{Hamiltonian}
\end{eqnarray}
where $\bS=(S_i)=\{+1,-1\}^N$ and 
$J_{i_1 i_2 \ldots i_p }$ is the quenched random coupling drawn from 
a Gaussian distribution of zero mean and the variance $p!/(2N^{p-1})$ for each ordered set of indices $( i_1 i_2 \ldots i_p )$. The conventional procedure using the TAP approach provides the TAP equation, as follows: 
\begin{eqnarray}
\tanh^{-1}(m_i)=\frac{\beta}{(p-1)!}
\sum_{j_2 j_3 \ldots j_p}
J_{i j_2 j_3 \ldots j_p} m_{j_2}m_{j_3} \ldots
m_{j_p}-m_i \frac{\beta^2p(p-1)}{2}(1-q)q^{p-2}, 
\label{TAPeq}
\end{eqnarray}
which represents the extremum condition of the TAP free energy
\begin{eqnarray}
&& \frac{1}{N}F_{ \rm TAP}(\bbm)=\frac{1}{N}\sum_{i=1}^N 
\left (
\frac{1+m_i}{2} \ln \frac{1+m_i}{2}
+\frac{1-m_i}{2} \ln \frac{1-m_i}{2} \right )
\cr
&& -\frac{\beta}{N}
\sum_{(i_1  i_2  \ldots  i_p)}
J_{i_1i_2 \ldots i_p} m_{i_1}m_{i_2} \ldots
m_{i_p}-\frac{\beta^2}{4}
((p-1)q^p-pq^{p-1}+1)
\label{TAPfree}
\end{eqnarray}
where $q=N^{-1}\sum_{i=1}^N m_i^2$. 
\subsection{BRST SUSY in Distribution of TAP Metastable States}
We here briefly summarize how the BRST SUSY concept is related to the 
TAP solutions following the argument presented in Ref. \cite{Crisanti2005}. 
For a specific sample of $\{J_{i_1i_2 \ldots i_p}\}$, 
the number of solutions that have a value of 
free energy density $f$ can be evaluated as
\begin{eqnarray}
\rho(f)&=&\sum_{\alpha}\prod_{i=1}^N \delta(m_i-m_i^\alpha)
\delta (F_{\rm TAP}(\bbm)-Nf) \cr
& = &\int d\bbm \prod_{i=1}^N
\delta (\partial_i F_{\rm TAP}(\bbm))
|\det (\partial_i \partial_j F_{\rm TAP}(\bbm))|
\delta (F_{\rm TAP}(\bbm)-Nf ), 
\label{dist_F_TAP}
\end{eqnarray}
where $\partial_i \equiv \partial/\partial m_i$ and 
$\alpha$ denotes the index of metastable states. 
In order to proceed further, 
we replace $|\det (\partial_i \partial_j F_{\rm TAP}(\bbm))|$ with 
$\det (\partial_i \partial_j F_{\rm TAP}(\bbm))$, expecting that the Hessian of most solutions contributing to the distribution is positive definite, 
and represent the determinant using fermionic variables
$\{\psi_i, \overline{\psi}_i\}$ as
\begin{eqnarray}
\det (\partial_i \partial_j F_{\rm TAP}(\bbm))
=\int {\cal D} \psi {\cal D} \overline{\psi }
\exp \left (
\sum_{ij} \overline{\psi}_i \psi_j
\partial_i \partial_j F_{\rm TAP}(\bbm)
\right ). 
\label{fermionic_det}
\end{eqnarray}
Using Eq. (\ref{TAPeq}) in the last delta function on $F_{\rm TAP}(\bbm)$ of Eq. (\ref{dist_F_TAP}) to eliminate the $J$-dependence in $F_{\rm TAP}(\bbm)$, which considerably reduces the work involved in the calculation but is valid only on the TAP solution, another expression of $\rho(f)$ is obtained, as follows:
\begin{eqnarray}
\rho(f)&=&
\int \cD \bx \cD \bbm \cD \bopsi \cD \bpsi
e^{\beta \cS(\bbm,\bx,\bopsi,\bpsi)}, 
\label{BRSTdist} \\
\cS(\bbm,\bx,\bopsi,\bpsi)
&=&\sum_{i}(x_i - \frac{u}{p}m_i) \partial_i F_{\rm TAP}(\bbm) \cr
&&+\sum_{ij} \opsi_i \psi_j \partial_i \partial_j F_{\rm TAP}(\bbm)
+u \left (F_{\rm TAP}(\bbm)
-Nf \right ), \label{BRSTaction}
\end{eqnarray}
where $\cD \ba \equiv \mbox{prefactor } \times \prod_i da_i$
and $u$ is determined so as to extremize the integral. 

The annealed distribution, which is considered in the present paper, is obtained by averaging Eq. (\ref{BRSTdist}) directly with respect to $\{J_{i_1i_2 \ldots i_p}\}$. The obtained result can be characterized by the annealed TAP complexity 
\begin{eqnarray}
\Sigma(f)&=&\frac{1}{N}\ln \overline{\rho}(f) \cr
&=& \mathop{\rm Extr}_{B,\Delta,\lambda,u}
\left \{
-fu \beta-(1-q)(B+\Delta)-q\lambda+
\frac{B^2-\Delta^2}{2q^{p-2}(p-1)\lambda \mu}+\ln I \right \},
\label{effect_action} 
\end{eqnarray}
where $\overline{\rho}(f)$ is the average of Eq. 
(\ref{dist_F_TAP}) with respect to $\{J_{i_1i_2 \ldots i_p}\}$, 
${\rm Extr}_a \{\cdots \}$ denotes the extremization of $\cdots$ 
with respect to $a$, $\mu=\beta^2p/2$, 
\begin{eqnarray}
I=\int_{-1}^1 
\frac{dm}{\sqrt{2 \pi \mu q^{p-1}}}
\left (\frac{1}{1-m^2} +B \right )
\exp \left [-\frac{(\tanh^{-1}(m)-\Delta m)^2}{2 \mu q^{p-1}}+
\lambda m^2+u \beta f(m) \right ], 
\label{macro_free}
\end{eqnarray}
and 
\begin{eqnarray}
\beta f(m)&=& \frac{1}{2} \ln (1-m^2)-\ln 2 +\frac{p-1}{p}m \tanh^{-1}(m)\cr
&&-\frac{\beta^2}{4}
[ 1+(p-2)q^{p-1}-(p-1)q^p ].
\label{onebodyfree}
\end{eqnarray}

Note that Eq. (\ref{BRSTaction}) is invariant under the generalized BRST transformation $\delta m_i=\epsilon \psi_i$; 
$\delta x_i= - \epsilon u (p-1)/p \psi_i$;
$\delta \opsi_i=-\epsilon x_i$;
$\delta \psi_i=0$, where $\epsilon$ is an infinitesimal fermionic 
constant. Such a property is referred to as the BRST SUSY. 
This guarantees that for any functions of 
$\{\bbm,\bx,\bopsi,\bpsi\}$, $\cO(\bbm,\bx,\bopsi,\bpsi)$, 
the average over the weight 
defined by action (\ref{BRSTaction}), 
$\left \langle \cO(\bbm,\bx,\bopsi,\bpsi) \right \rangle $, 
is also invariant under the BRST transformation, 
which can be used to yield various identities. 
In particular, two identities 
\begin{eqnarray}
\LL \opsi_i \psi_i \RR &=&  -\LL m_i x_i \RR+\frac{u}{p} \LL m_i^2 \RR, 
\label{BRST1}\\
u \LL \opsi_i \psi_i \RR &=&  \LL x_i^2 \RR-\frac{2u}{p}\LL m_i x_i \RR +\frac{u^2}{p^2}
\LL m_i^2 \RR,  
\label{BRST2}
\end{eqnarray}
which are obtained by setting $\cO=m_i \opsi_i$ and $\cO=x_i \opsi_i$, respectively, 
which is useful for solving the rather complicated extremization problem of Eq. (\ref{effect_action}). Specifically, by imposing two conditions $\Delta=-\beta^2(p-1)uq^{p-1}/2$ and $\lambda=\beta^2(p-1)^2u^2 q^{p-1}/(4p)$, which represent two relations that are obtained by averaging the BRST SUSY identities of Eqs. (\ref{BRST1}) and (\ref{BRST2}) over $\{J_{i_1i_2 \ldots i_p}\}$ with appropriate weights, on Eq. (\ref{effect_action}) and setting $B=0$ following BM, a considerably simplified expression of the TAP complexity is obtained \cite{Cavagna2003}, as follows:
\begin{eqnarray}
\Sigma_{\rm BRST}(f)=\mathop{\rm Extr}_{q,u} 
\left \{ \beta u [F_{\rm 1RSB}(\beta;q,-u)-f ] \right \}. 
\label{BRST_sol}
\end{eqnarray}
Here, 
\begin{eqnarray}
\beta F_{\rm 1RSB}(\beta;q_1,x)
=-\frac{1}{x}\ln \int Dz 
\left (2 \cosh \left (\beta \sqrt{\frac{p}{2}q_1^{p-1}} z \right )\right )^x
-\frac{(1-x)}{4} \beta^2J^2q_1^p-\frac{\beta^2J^2}{4}, 
\label{1RSB_free}
\end{eqnarray}
where $Dz=dz \exp [-z^2/2]/\sqrt{2 \pi}$, indicates the one-step RSB (1RSB) variational free energy of the current $p$-body spin glass systems, the extremization 
of which provides the values of self-overlap $q_1$ and Parisi's RSB parameter $x$ in the equilibrium state. 

In addition to Eq. (\ref{BRST_sol}), the extremization problem of Eq. (\ref{effect_action}) yields another solution, which does not satisfy the BRST SUSY relations. This solution, which we refer to as $\Sigma_{\rm BM}(f)$, agrees with that which was previously discovered by the conventional recipe of BM \cite{BrayMoore1980}. For $p=2$, $\Sigma_{\rm BM}(f)$ always dominates $\Sigma_{\rm BRST}(f)$ and, therefore, the BRST SUSY solution (\ref{BRST_sol}) is not physically relevant, which is also supported by the numerical experiments \cite{Cavagna2004}. 
However, for $p > 2$, in particular, $p=3$, $\Sigma_{\rm BRST}(f)$ overcomes $\Sigma_{\rm BM}(f)$ in a lower free energy region, implying the physical relevance of the present approach. We herein focus primarily on the justification of this theoretical prediction by extensive numerical experiments. 

\section{Numerical Experiment}
\subsection{Numerical Scheme}
Computational cost is a major difficulty in numerically solving the TAP equation for $p \ge 3$. The cost for assessing the right-hand side of Eq. (\ref{TAPeq}) increases $O(N^{p-1})$ per element, which practically reduces the upper limit size of systems that can be numerically examined as $p$ grows. Since extensive experiments have only been performed up to $N=80$, even for the case of $p=2$, reducing the computational cost to the greatest extent possible is indispensable. 

Solving the TAP equation by natural iteration of Eq. (\ref{TAPeq}) is not possible \cite{TAPdifficulty}. Instead, gradient descent schemes of a master function $L(\bbm)=\sum_{i=1}^N \left (\partial_i F_{\rm TAP}(\bbm) \right )^2$, by which convergence to a certain state is guaranteed, have frequently been employed \cite{NemotoTakayama,DewarMottishaw1988,Nemoto1988,NishimuraNemoto1990}. However, simple gradient methods generally require a number of iterations as the locally determined direction of steepest descent does not necessarily approximate well the direction to the nearest local minimum, which requires several updates for convergence to the local minimum. A solution to this difficulty is the use of a quasi Newton method, which was used in the investigation of $p=2$ systems \cite{Cavagna2004}. Unfortunately, the use of this scheme is also difficult for our purpose because assessment of the Jacobian required in this algorithm 
has a higher computational cost than that at the practically feasible level, even if the efficient formula proposed by Sherman-Morrison 
is used \cite{NumRec}. 

In order to avoid such difficulties, we employed a simple discrete algorithm that guarantees a reduction in the TAP free energy at each update as follows. The diagonal elements of the Hessian of the TAP free energy are provided as follows:
\begin{eqnarray}
\partial_i^2 F_{\rm TAP}(\bbm)=
\frac{1}{1-m^2_i}+\frac{\beta^2p(p-1)}{2}(1-q)q^{p-2} \ge 0. 
\label{diag_hessian}
\end{eqnarray}
This indicates that solving Eq. (\ref{TAPeq}) with respect to a single variable $m_i$ while maintaining the other variables $\{m_{j\ne i}\}$ constant yields a unique solution, which can be searched by an efficient scheme such as the bisection method. In addition, updating $m_i$ to the solution guarantees reduction of the TAP free energy (\ref{TAPfree}). This implies that for an arbitrary initial state, 
iteration of the above procedure leads to convergence of a certain {\em local minimum} of the TAP free energy (\ref{TAPfree}). Hopefully, applying this algorithm to a sufficient number of initial states while randomly permuting the order of updates among elements will yield a complete set of local minima for a given fixed set of $\{J_{i_1i_2\ldots i_p}\}$. 

A major drawback of this strategy may be that extrema of other types, namely, saddle points and maxima cannot be obtained, although such solutions are taken into account 
in Eq. (\ref{dist_F_TAP}). If the contribution of saddles and maxima to the complexity are relevant, this implies that the proposed strategy is not appropriate for verifying Eq. (\ref{dist_F_TAP}) or Eq. (\ref{effect_action}). However, recent research into the case of $p=2$ revealed that most solutions of the TAP equation consist of a pair of a local minimum and a saddle point, for which only one direction 
has a negative eigenvalue \cite{Aspelmeier2003,Cavagna2004}. This implies that the complexity composed of only local minima yields the same result as that evaluated from all extrema, justifying a crucial approximation $|\det (\partial_i \partial_j F_{\rm TAP}(\bbm)) | \simeq \det (\partial_i \partial_j F_{\rm TAP}(\bbm))$ a posteriori. We, therefore, employ the minimization algorithm 
with the expectation that this property also holds for cases in which $p \ge 3$, 
which should be investigated in the future. 

\subsection{Comparison with Theoretical Predictions}
\begin{figure}
\setlength{\unitlength}{1mm}
\includegraphics[width=140mm]{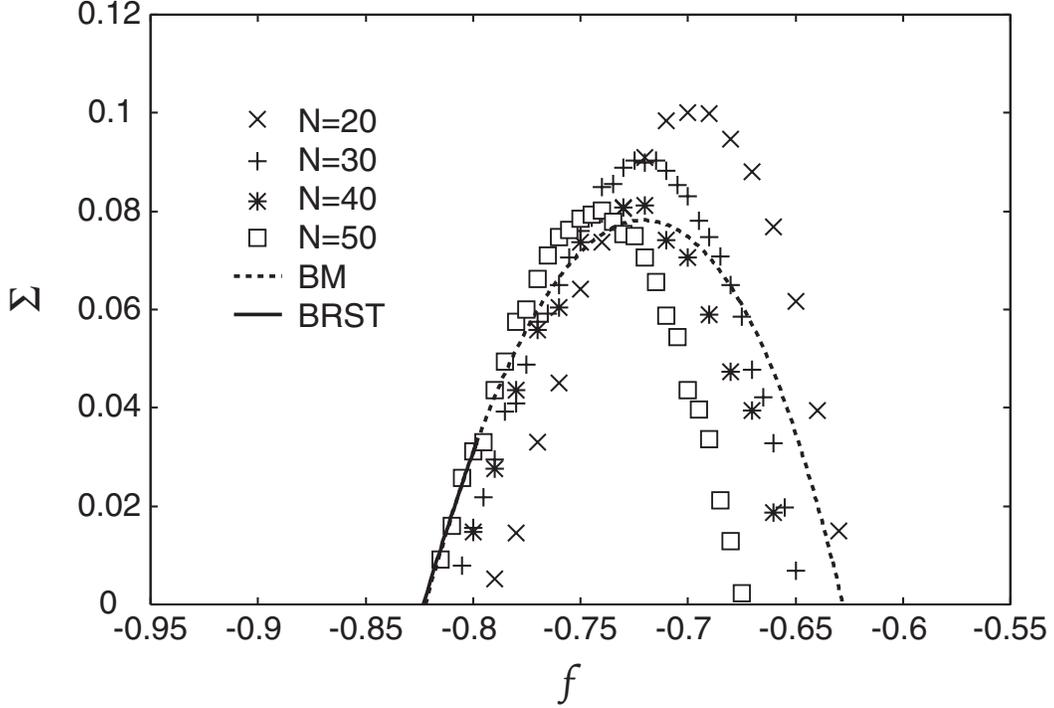}
\caption{Annealed complexity vs. free energy density for various values of $N$.
The solid and dotted lines represent 
$\Sigma_{\rm BRST}(f)$ and $\Sigma_{\rm BM}(f)$, respectively. 
$\Sigma_{\rm BRST}(f)$ agrees with $\Sigma_{\rm BM}(f)$
and is terminated at critical point $f_c=-0.800\ldots$. }
\label{complexity}
\end{figure}

We performed numerical experiments of $p=3$ systems for $N=20, 30$, $40$ and $50$ at $T=T_s=0.5$, which corresponds to the region of 1RSB in equilibrium analysis. 
The number of samples of $\{J_{i_1 i_2 \ldots i_p}\}$ was $25800$ for $N=20$, $17200$ for $N=30$, $1720$ for $N=40$ and $430$ for $N=50$. Unfortunately, it was difficult to efficiently find solutions other than the trivial paramagnetic solution $\bbm=\bzero$ from randomly chosen initial states, which requires a practically infeasible number of trials. In order to raise the efficiency of finding non-trivial solutions, we first solved the naive mean field equation 
$\tanh^{-1}(m_i)=\frac{\beta}{(p-1)!}
\sum_{j_2 j_3 \ldots j_p}
J_{i j_2 j_3 \ldots j_p} m_{j_2}m_{j_3} \ldots
m_{j_p}$ of $T=T_s$ from many initial states, which can efficiently produce a number of non-trivial solutions. The obtained solutions are employed as initial states for solving Eq. (\ref{TAPeq}).
For finding solutions uniformly, 
we randomly permuted the order of updates among elements
at each iteration indexed by $t$; the convergence was judged by 
a condition $\sum_{i=1}^N |m_i^{t}-m_i^{t-1}| \le 10^{-8}$. 
In order to ensure that our search of solutions is as exhaustive as possible, we repeated the search $500$ times for $N=20$, $500$ times for $N=30$, $2000$ times for $N=40$ and $5000$ times for $N=50$, respectively, for a given sample of connections $\{J_{i_1 i_2 \ldots i_p}\}$. All solutions found in samples were found at least five times. 
The Hessian was positive definite for all the solutions, 
which indicates that our simple scheme successfully worked as a 
minimization algorithm. 

Based on a previous study \cite{Cavagna2004}, we numerically assessed the complexity by a formula $\Sigma(f)=\ln \left [p(f) {\cal N} \right ]/N$, where $p(f)$ is the probability of finding a solution of the free energy density $f$ among the total set of numerically found solutions and ${\cal N}$ is the average number of solutions for one sample set of connectivities. Theoretical assessment predicts that the BRST SUSY solution $\Sigma_{\rm BRST}(f)$ slightly dominates the BM solution $\Sigma_{\rm BM}(f)$ in the region of lower free energies $f < f_c = -0.800 \ldots$, while $\Sigma_{\rm BM}(f)$ is dominant for $f > f_c$. 

\begin{figure}
\setlength{\unitlength}{1mm}
\includegraphics[width=140mm]{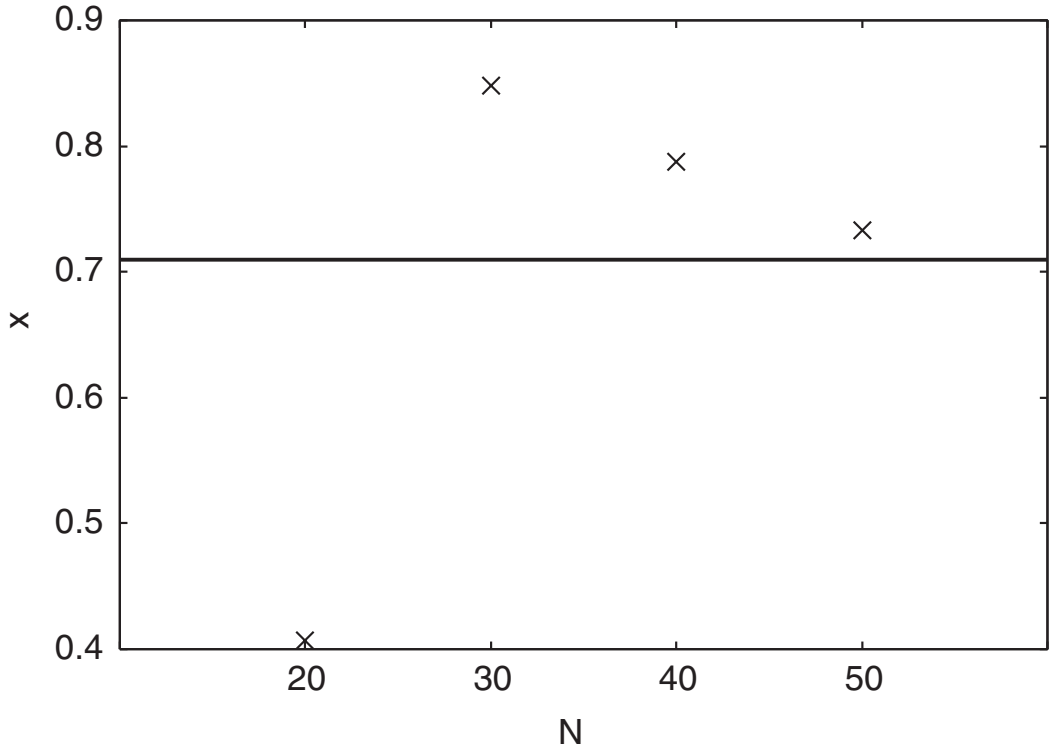}
\caption{Parisi's 1RSB parameter $x$ vs. $N$. The crosshairs denote numerical estimates assessed from slope at the left terminal of the complexity $f_l$ for which 
$\Sigma(f_l;N)=0$ by quadratic fitting. The line denotes the theoretical
value for $T=T_s=0.5$, $x=0.710\ldots$. }
\label{parisi_param}
\end{figure}

Figures \ref{complexity} show a comparison of complexity between the theoretical predictions and numerical data. These exhibit reasonable consistency in the lower free energy region, where $\Sigma_{\rm BRST}(f)$ is dominant, except for $N=20$, while the discrepancy is considerable for larger $f$. Solutions of high free energy have been reported to be difficult to obtain \cite{Cavagna2004,Nemoto1988,NishimuraNemoto1990,Plefka2002}, which is thought to be the reason why in the present study the complexity of larger $f$ is underestimated as $N$ increases. Based on the formal equivalence between the BRST SUSY solution of the TAP framework and the 1RSB solution in the replica analysis, the tangent slope at the left terminal of the complexity $f_l=-0.823\ldots$ for which $\Sigma_{\rm BRST}(f_l)=0$ is expressed as $\beta x$, where $x$ is the Parisi's 1RSB parameter, the value of which is determined by the extremization condition of the 1RSB free energy (\ref{1RSB_free}). 
Figure \ref{parisi_param} shows that numerical estimates of $x$ assessed by the quadratic fitting of the complexity is in relatively good agreement with the theoretical value of $x=0.710\ldots$ for $T=T_s=0.5$ except for $N=20$. 

\begin{figure}
\setlength{\unitlength}{1mm}
\includegraphics[width=140mm]{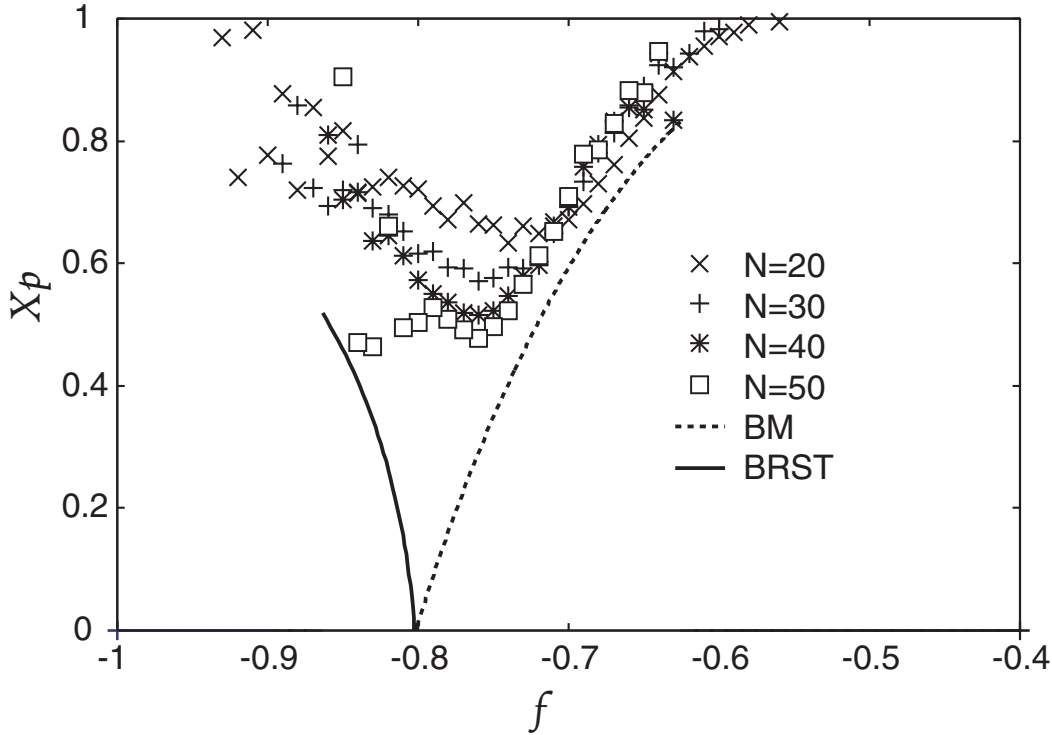}
\caption{Plefka's stability parameter vs. free energy density for various values of $N$. The solid and dotted curves represent the values assessed 
by the BRST SUSY and BM solutions, respectively. }
\label{x_p}
\end{figure}

Tempering local minima while maintaining the relation 
(\ref{TAPeq}) from $T=0$ to $T_s$ might be another scheme
for improving the performance of the solution search \cite{NemotoTakayama}. 
However, such a strategy was found to be not so effective, particularly for 
finding solutions of {\em lower free energies}. At $T=0$, the complexity of this system is expressed by the BM solution of BRST SUSY breaking in the entire region of $f$. Therefore, it can be expected that solutions of $T=T_s$, which are adiabatically linked to those of $T=0$, are mostly characterized as BRST SUSY breaking unless the properties of the solutions are changed. This may be the reason why solutions of lower free energy at $T=T_s$, which are characterized by the BRST SUSY complexity, are difficult to find by the tempering scheme. 
\begin{figure}
\setlength{\unitlength}{1mm}
\includegraphics[width=140mm]{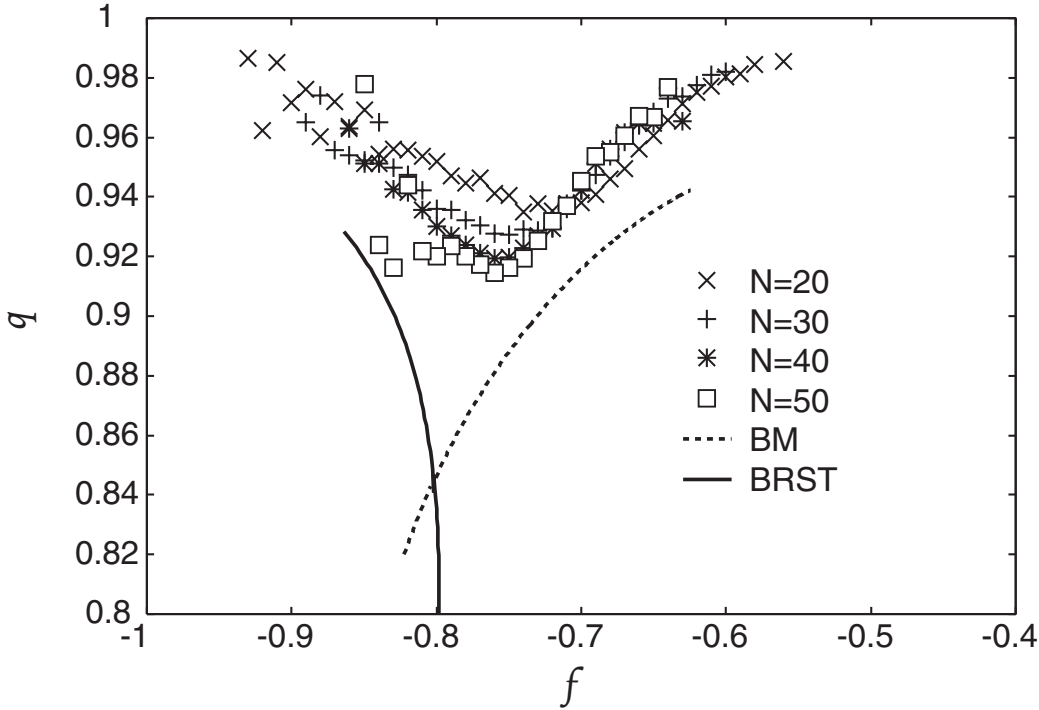}
\caption{Self-overlap vs. free energy density for various values of $N$.
The solid and dotted curves represent values assessed 
by the BRST SUSY and BM solutions, respectively. }
\label{overlap}
\end{figure}

It is difficult to observe the transition at $f=f_c$ between the BRST SUSY and BM solutions by numerically assessing the complexity because the difference between $\Sigma_{\rm BRST}(f)$ and $\Sigma_{\rm BM}(f)$ is considerably small, even in the region where $\Sigma_{\rm BRST}(f)$ dominates $\Sigma_{\rm BM}(f)$. However, this is not the case for other quantities. 

Figure \ref{x_p} shows a plot of Plefka's stability parameter \cite{Plefka1982}
\begin{eqnarray}
x_p=1-\frac{\beta^2p (p-1)}{2} \left \langle (1-m^2)^2 \right \rangle, 
\label{plefka}
\end{eqnarray}
versus free energy density $f$. Theoretical assessment indicates that $x_p$ vanishes
to zero at $f=f_c$ signaling the BRST SUSY-SUSY breaking transition, while it is maintained positive at other locations (curves). Although the deviation from the theoretically obtained curves is relatively large due to the finite size effect, the numerical data are minimized in the vicinity of $f=f_c$, which is qualitatively in accordance with the above mentioned behavior. Similar behavior is also observed for self-overlap $q$ (Figure \ref{overlap}). 

\section{Summary}
In summary, we have numerically examined the properties of the TAP solutions for $p$-body Ising spin glass systems, in particular, for $p=3$ cases. A theoretical study reported that the complexity of this system is expressed by the BRST SUSY solution in the lower free energy region when the equilibrium state is expressed by the 1RSB solution, while the BM solution of BRST SUSY breaking is relevant for the higher free energy region \cite{Crisanti2005}. As another study concluded that the BRST SUSY solution is always irrelevant for $p=2$ cases \cite{Cavagna2004}, the emergence of the BRST SUSY solution in the lower free energy region is a discriminative feature of $p(\ge 3)$-body systems. 

In order to justify this feature, we have performed extensive numerical experiments employing a simple discrete algorithm, which locally minimizes the TAP free energy (\ref{TAPfree}). A heavy computational cost practically limits the experiments for systems of relatively small sizes of less than $N=50$, which makes it difficult to quantitatively distinguish the BRST SUSY solution from that of BM in complexity curves by numerical methods. However, other quantities, such as the Plefka's stability parameter $x_p$ and self-overlap $q$, are expected to exhibit peculiar behavior as a function of free energy density $f$ as a consequence of the BRST SUSY-SUSY breaking transition, which has been qualitatively reproduced by our numerical experiments. This may indicate that the TAP solutions of lower free energy of $p(\ge 3)$-body systems in 1RSB phase, which govern equilibrium properties, are characterized as having BRST SUSY, unlike $p=2$ systems. 

Application of the current approach to sparsely connected systems is left as a topic for future study \cite{DewarMottishaw1988}. 

\begin{acknowledgments}
The authors would like to thank Prof. Isao Ono for the use of the DIS-Cluster System of Department of Computer Intelligence and Systems Science, Tokyo Institute of Technology. This work was supported in part by Grants-in-Aid from JSPS/MEXT, Japan, (Nos. 17340116 and 18079006). 
\end{acknowledgments}

\end{document}

%% file: commands.tex
\newcommand{\sign}{\mbox{\rm{sign}}}
\newcommand{\opsi}{\overline{\psi}}
\newcommand{\btau}{\mbox{\boldmath{$\tau$}}}
\newcommand{\brmP}{\mbox{\bf{P}}}
\newcommand{\brmI}{\mbox{\bf{I}}}
\newcommand{\brmC}{\mbox{\bf{C}}}
\newcommand{\brmW}{\mbox{\bf W}}
\newcommand{\bU}{\mbox{\boldmath{$U$}}}
\newcommand{\blambda}{\mbox{\boldmath{$\lambda$}}}
\newcommand{\bgamma}{\mbox{\boldmath{$\gamma$}}}
\newcommand{\bn}{\mbox{\boldmath{$n$}}}
\newcommand{\bh}{\mbox{\boldmath{$h$}}}
\newcommand{\bH}{\mbox{\boldmath{$H$}}}
\newcommand{\bX}{\mbox{\boldmath{$X$}}}
\newcommand{\bD}{\mbox{\boldmath{$D$}}}
\newcommand{\bb}{\mbox{\boldmath{$b$}}}
\newcommand{\br}{\mbox{\boldmath{$r$}}}
\newcommand{\bbs}{\mbox{\boldmath{$s$}}}
\newcommand{\bS}{\mbox{\boldmath{$S$}}}
\newcommand{\bone}{\mbox{\boldmath{$1$}}}
\newcommand{\bxi}{\mbox{\boldmath{$\xi$}}}
\newcommand{\bv}{\mbox{\boldmath{$w$}}}
\newcommand{\bw}{\mbox{\boldmath{$w$}}}
\newcommand{\bx}{\mbox{\boldmath{$x$}}}
\newcommand{\bbm}{\mbox{\boldmath{$m$}}}
\newcommand{\bpsi}{\mbox{\boldmath{$\psi$}}}
\newcommand{\bopsi}{\mbox{\boldmath{$\opsi$}}}
\newcommand{\bM}{\mbox{\boldmath{$M$}}}
\newcommand{\bL}{\mbox{\boldmath{$L$}}}
\newcommand{\bA}{\mbox{\boldmath{$A$}}}
\newcommand{\ba}{\mbox{\boldmath{$a$}}}
\newcommand{\be}{\mbox{\boldmath{$e$}}}
\newcommand{\bR}{\mbox{\boldmath{$R$}}}
\newcommand{\bO}{\mbox{\boldmath{$O$}}}
\newcommand{\by}{\mbox{\boldmath{$y$}}}
\newcommand{\bz}{\mbox{\boldmath{$z$}}}
\newcommand{\bzero}{\mbox{\boldmath{$0$}}}
\newcommand{\bzeta}{\mbox{\boldmath{$\zeta$}}}
\newcommand{\bJ}{\mbox{\boldmath{$J$}}}
\newcommand{\cD}{{\cal D}}
\newcommand{\cJ}{{\cal J}}
\newcommand{\cS}{{\cal S}}
\newcommand{\cG}{{\cal G}}
\newcommand{\cO}{{\cal O}}
\newcommand{\LL}{\left \langle}
\newcommand{\RR}{\right \rangle}